\documentclass[fleqn,10pt]{wlscirep}
\usepackage[utf8]{inputenc}
\usepackage[T1]{fontenc}
\usepackage[official]{eurosym}
\usepackage{setspace}

\newcommand{\COT}{CO$_2$}
\newcommand{\tceq}{tCO$_2$eq}

\title{An astronomical institute's perspective on meeting the
  challenges of the climate crisis}

\author[1,*,+]{Knud Jahnke}
\author[1,+]{Christian Fendt}
\author[1,+]{Morgan Fouesneau}
\author[1,+]{Iskren Georgiev}
\author[1,+]{Tom Herbst}
\author[1,+]{Melanie Kaasinen}
\author[1,+]{Diana Kossakowski}
\author[1,+]{Jan Rybizki}
\author[1,+]{Martin Schlecker}
\author[1,+]{Gregor Seidel}
\author[1]{Thomas Henning}
\author[1]{Laura Kreidberg}
\author[1]{Hans-Walter Rix}

\affil[1]{Max Planck Institute for Astronomy, Heidelberg, Germany}
\affil[*]{jahnke@mpia.de}
\affil[+]{these authors contributed equally to this work}

\begin{abstract}
Analysing greenhouse gas emissions of an astronomical institute is a
first step in reducing its environmental impact. Here, we break down
the emissions of the Max Planck Institute for Astronomy in Heidelberg
and propose measures for reductions.
\end{abstract}

\begin{document}

\flushbottom
\maketitle


Humanity's production of greenhouse gas (GHG) emissions is threatening
our own habitat, our physical and mental health, and the chances of
long-term survival of human society as we know it
\cite{WHO_2003,IPCC_policymakers}.  The greenhouse gases emitted as we
burn fossil fuels for energy have already resulted in a mean surface
temperature rise of more than 1$^\circ$\,C since the late 19th
century\cite{berkeleyearth2019}. To further limit the temperature rise
to less than 1.5$^\circ$\,C as per the Paris
agreement\cite{parisagreement} requires all sections of human society
to reduce their GHG emissions to net zero by 2050. The scientific
profession is not exempt. It is our responsibility to analyse the
origin of our work-related emissions, to identify solutions for
reducing emissions, and to determine the responsibility on a personal,
institute-, community-, and society-wide level for implementing the
necessary changes.

As astronomers of the Max Planck Institute for Astronomy (MPIA) in
Heidelberg, Germany, we have assessed our work-related GHG
emissions. The MPIA is a well-funded, international astronomy research
institute with $\sim $150 researchers and $\sim$320 employees in
total. A wide range of research is conducted at the institute,
including the development of astronomical instrumentation, analysis of
observational data, and theoretical modeling of astrophysical
phenomena with computing facilities. The institute is scientifically
well connected both within Europe and internationally, which, in
combination with the broad range of research departments, makes it a
good test case for the analysis of research-associated GHG
emissions. This report can therefore serve as a template for other
institutes. Our analysis provides a complementary, European
perspective to the analysis by the Australian astronomical
community\cite{ausastr}, the Canada France Hawaii Telescope
\cite{cfht2020}, the annual European Astronomical Society conference
\cite{ewass2019}, and an earlier analysis of US
astronomy\cite{marshall2009}.

\section*{MPIA greenhouse gas emissions}

We assessed the MPIA's GHG emissions in seven categories; business
flights, commuting, electricity, heating, computer purchases, paper
use, and cafeteria meat consumption. These categories were selected
either because they were likely to have a large contribution or
because we had no prior gauge of their significance.  For this first
assessment, we omitted other purchases, including materials and
components for instrumentation, additional office supplies, and IT
hardware other than desktop and laptop computers.

The GHG emissions associated with some categories were easily
determined, for example from electricity and heating oil bills,
computer expenses, and paper purchases and recycling amounts. However,
other categories proved less straightforward. Assessing the emission
from flights required both a manual transcription of invoices and a
questionnaire to all employees about self-booked business trips, as
there was no automated and accessible list of itineraries, carriers or
classes. Nevertheless, all numbers quoted here (see
Table~\ref{tab:emissions}) are capturing the MPIA's 2018 emissions
quite well.  We estimate the major contributors to our greenhouse gas
emissions, that is, flying and electricity, to be accurate to within
20\%.

Table~\ref{tab:emissions} summarizes the emission sources and the
associated quantities. We have converted the units for each source
into tons of \COT-equivalent emissions (\tceq). The term
``equivalent'' denotes that these values are normalized to the GHG
impact of \COT. In particular, the numbers in this table account for
flight emissions at altitude (e.g.\ soot, sulphates, nitrogen oxides, and
cirrus clouds from contrails), as well as methane emissions from meat
farming.

\begin{table}[ht]
\centering
\begin{tabular}{|l|c|r|r|r|}
\hline
{\bfseries Source} & {\bfseries Amount} & {\bfseries \COT eq} & {\bfseries \COT eq/researcher}&{\bfseries Percentage (\%)}\\
\hline
\hline
Travel (air)                &1030 flights       & 1280\,t  &8.5\,t& 47\\
\hline
Electricity (on/off campus) &3,400,000 kWh      & 779\,t   &5.2\,t& 29\\
\hline
Heating (oil)               &150,000 l          & 446\,t   &3.0\,t& 16\\
\hline
Commuting (car)             &792,000 km         & 139\,t   &0.9\,t& 5\\
\hline
Paper / cardboard           & 0.15 / 7 t        & 35\,t    &0.2\,t& 1\\
\hline
Computer (desk-/laptops)    & 57 purchased          & 29\,t    &0.2\,t& 1\\
\hline
Meat (canteen)              & 1000 kg           & 16\,t    &0.1\,t& $<$1\\
\hline
\hline
{\bfseries Total}                       &                  & {\bfseries $\sim$2720} &{\bfseries 18.1\,t}& 100\,\%\\
\hline
\end{tabular}
\caption{
\label{tab:emissions}
Summary of the MPIA's GHG emissions in 2018. Note that
electricity includes both consumption at the MPIA campus, as well as
in external supercomputing centers used by MPIA.}
\end{table}

The MPIA's total GHG emissions for 2018 amount to 18.1\,\tceq\ per
researcher. Alternatively, the contribution per refereed science
publication, of which there were 583 either authored or co-authored by
MPIA astronomers in 2018, is 4.6\,\tceq. However, regardless of the
chosen denominator, these metrics have caveats in attribution. For
example a substantial part of the institute's emissions results from
instrumentation projects that will lead to future publications but at
the same time, we also do not account for the emissions associated
with the construction of observing facilities used in the 2018 papers;
also simulations can take months to years.

\begin{figure}[ht]
\centering
\includegraphics[width=0.8\linewidth]{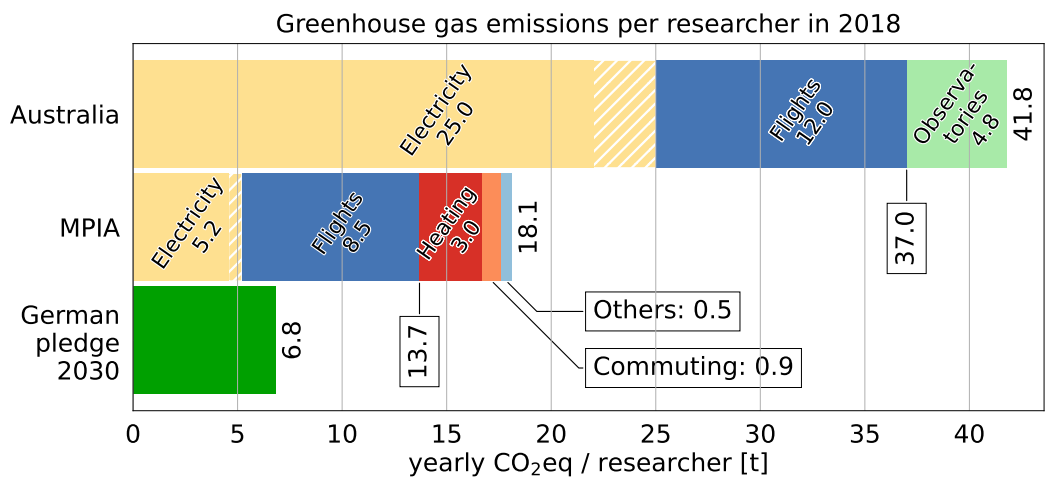}
\caption{Average annual emissions in 2018 for an Australian and MPIA
  researcher in \tceq/yr, broken down by sources. The sources include
  electricity, flights (converted to the same emission model, see
  text), observatory operation, office heating, commuting, and
  `others', a category that combines office desktop and laptop
  hardware, paper and cardboard use, and meat consumption. Electricity
  related emissions include both computing and non-computing
  consumption, where for Australia computing is accounting for 88\% of
  electricity emissions; we estimate a similar fraction for MPIA. In
  the plot, the smaller hatched part of the 'Electricity' bar
  indicates non-computing electrical power. Observatory operation is
  only given for Australia, while heating, commuting, and sources
  captured by the `others' category are only given for
  MPIA. Therefore, emissions can only be compared between Australia
  and MPIA for electric power consumption and flights, which amount to
  37.0 and 13.7 \tceq/yr for Australian and MPIA researchers
  respectively. The major difference lies in the amount of GHG
  emissions per kWh electricity, which differs by a factor of $\sim$4
  between the Australian astronomy and the MPIA.  These values do not
  account for all emissions per capita. In particular, emissions not
  related to work are excluded. The combined MPIA emissions of 18.1
  \tceq/yr and researcher are also compared to the German pledge of a
  55\% reduction of the 1990 emissions by 2030, plotted per capita in
  dark green, which is close to 6.8 \tceq/capita per year
  \cite{bundesklimaschutzgesetz,germanCO2report,destatis1990}.  }
\label{fig:average_emission_by_source_and_place}
\end{figure}

The MPIA's astronomy-related GHG emissions per researcher in 2018 were
an alarming $\sim$3 times higher than the German target for 2030 (in
line with the Paris Agreement; see
Figure~\ref{fig:average_emission_by_source_and_place})
\cite{germanCO2report,bundesklimaschutzgesetz,destatis1990}. Moreover,
the per-researcher emissions are $\sim$60\% higher than for the
average German resident, whose annual 2018 GHG emissions (by
consumption) were
11.6\,\tceq\cite{germanCO2report,destatis,owidco2andothergreenhousegasemissions}
(GHG emissions by consumption per adult resident were
14.0\,\tceq\cite{destatis}).  Of course, these numbers just compare
the work-related contributions of MPIA researchers to the Paris target
and German averages, neglecting the additional emissions associated
with non-research related ``private'' emissions by MPIA researchers,
as for example housing, clothing, private mobility, or food.

Few comparisons exist in the astronomical context. We therefore
compare the MPIA's emissions to the assessment by the Australian
astronomical community \cite{ausastr}. The MPIA's per astronomer
emissions are approximately half that of the Australian astronomer,
which amount to 42\,\tceq\ per capita (see
Figure~\ref{fig:average_emission_by_source_and_place}). Note that we
calculated flight emissions using the model by
atmosfair.de\cite{atmosfair}, which estimates approximately double the
emissions of the Qantas calculator \cite{qantas} used for the original
Australian assessment\cite{ausastr}.  Adjusting the reported
Australian number by this factor, the MPIA's flight emissions are
similar or somewhat lower than that for the Australian astronomical
community. The second major contributor to the MPIA's GHG emissions is
our electricity consumption, at $\sim$5\,\tceq\ per astronomer. In
contrast, the electricity-related emission, at 22\,\tceq\ per
astronomer, dominated the Australian astronomer's GHG emissions. The
MPIA's electricity consumption mainly results from our computing
needs, which for 2018 also included the use of supercomputing
facilities in Garching (Max Planck Computing and Data Facility), and
at the University of Stuttgart for a specific large-scale simulation
project. However, the difference to Australia in electricity-related
emissions is almost completely due to the different carbon intensity
for electricity production: Whereas fossil fuel sources contributed
83\% to Australia's generation of electricity in 2018
\cite{Aus_energy_stats}, the contribution in Germany was
$\sim$47\%\cite{fraunhoferISE}, and MPIA's delivery contracts have a
carbon intensity even substantially below that.  Thus, for the
Australian, community the electricity usage for computing is
calculated to require 0.905\,kg\COT/kWh, whereas MPIA's electricity
contracts average 0.23\,kg\COT/kWh.  Lastly, we note that the MPIA's
heating oil emissions in 2018 are comparable to the ``campus
operation" emissions derived for the Australian community (both
3\,\tceq\ per researcher), which are extrapolated from the building
power requirements of one institute.

\medskip

\section*{Potential measures to reduce emissions}

To reduce our astronomy-related GHG emissions, we need to identify
which measures will be effective and need implementation at which
level, i.e.\ at the level of the individual researchers, the MPIA, the
Max Planck Society, the astronomical community, or human society in
general. Each institute will face its specific challenges. For
example, we have identified the high carbon intensity of MPIA's
heating, which needs to be addressed at the institute level, but other
measures need changes across the astronomical community. Measures and
responsibilities can only be identified once the GHG emissions have
been quantified.

\subsection*{Flying}

Flight-related GHG emissions dominate the MPIA's total
emissions. Since there is no technology on the horizon that would
reduce flight emissions to anything approaching carbon-neutral by
2050, much less 2030, the only way to reduce flight-related emissions
is to reduce this form of travel. To do so, we need to identify the
destinations and reasons for the air travel.

\begin{figure}[ht]
\centering
\includegraphics[width=0.8\linewidth]{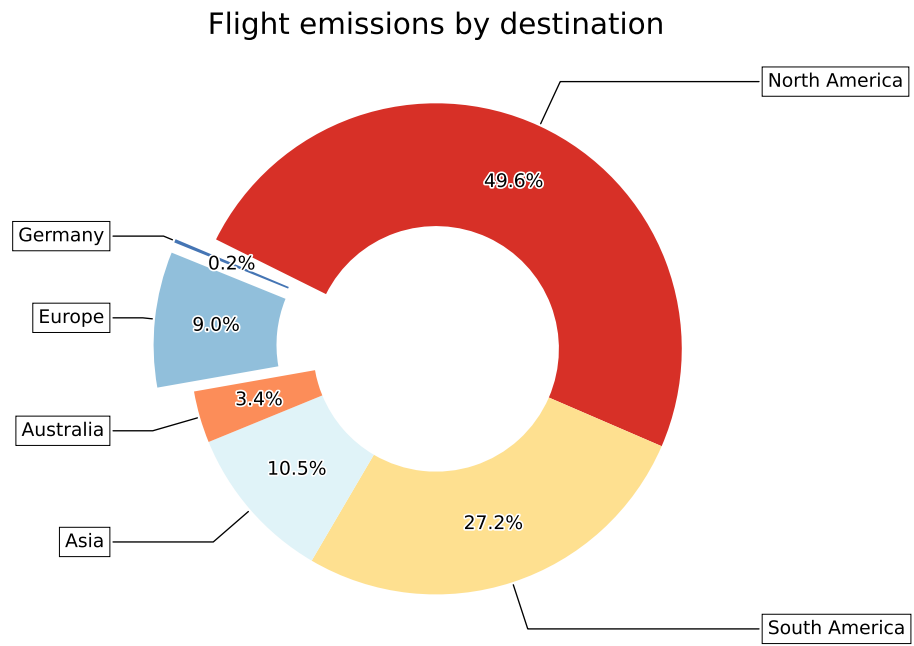}
\caption{Relative GHG emissions broken down by flight destination for
  MPIA employees. Intercontinental flights that cannot be easily
  replaced by alternative means of transport make up about 91\% of
  flying emissions. This is due to the number of flights, and the high
  climate impact of each intercontinental flight, primarily due to
  distance traversed, but also due to greater time averaged emission
  altitude, for example for nitrogen oxides.}
\label{fig:flight_emissions_by_destination}
\end{figure}

In Figure~\ref{fig:flight_emissions_by_destination}, we break down the
MPIA's emissions by destination. A negligible fraction of emissions
originates from flights inside Germany, and only 9\% from flights with
destinations inside Europe (including the Canary Islands). Though
small, this European component can be further reduced by replacing air
with train travel\cite{ewass2019}. Changes to the German public
servant's travel law in early 2020 ensure that train trips to
well-connected European destinations are now reimbursed, even if they
are more expensive than a flight\cite{brkg}. Moreover, at the
individual level, many German researchers have pledged not to fly
distances under 1000\,km\cite{SFF1000}. However, the vast majority of
the MPIA's flight emissions ($>$90\%) stem from intercontinental
flights, which are dominated by destinations in the USA and
Chile. Although we cannot identify the reason for each flight, in
general, these international flights are a mix of travel for
observation campaigns, instrument commissioning, conferences,
seminars, and research visits. To reduce our flight-related emissions,
we must identify solutions that enable us to reach the scientific
goals of these trips without the need for air travel. The onus here is
on the entire astronomical community to change how we work.

Travel for meetings, conferences and collaborations made up a
significant fraction of the MPIA's 2018 flights, as will be the case
for most astronomical institutes. At that time, video-based
alternatives were only used in specific settings. However, the need to
continue working during the 2020 Covid-19 pandemic resulted in the
substitution of many physical meetings with virtual ones.  To reduce
our carbon footprint to anything approaching net zero by 2050, the
expertise in hosting virtual events that was so rapidly developed
during the last few months, should continue to be applied and
expanded. To this end, the recommendations of Klöwer et
al.\ \cite{kloewer2020} are an excellent starting point. They are
providing an in-depth analysis of conferencing carbon emissions and an
overview of options. Their analysis shows that GHG emissions for
in-person meetings will strongly depend on the meeting location
relative to the origin of the participants, and they make cases
e.g.\ for fully online meetings and hybrid models with continental
in-person meeting ``hubs'', combined with online connections between
hubs, as well as other changes that would drastically reduce the
conference-induced emissions. These and other models in combination
with a drastically lower number of conferences promise to be an
effective measure.

In contrast, we identify reasons for flights for which we have no
immediate alternatives. These include, for example, extended in-person
collaborative visits, that prove very effective for initiating new
projects, and the installation or commissioning of instruments at
telescopes including the LBT (Arizona) or the ESO VLT
(Chile). Hardware built by the MPIA must be mounted at a telescope
site, tuned, and put into science operations, and as a result expert
engineers and astronomers have to be physically present for a larger
number of commissioning runs. Hypothetically, some runs could be
combined, but this immediately impacts engineering timescales and
family boundary conditions that might be complex to solve. The
institute and the astro community have to search for measures to
address these cases, which at this point are unsolved.

\subsection*{Computing-related Emissions}

The second major contributor to the MPIA's GHG emissions resulted from
the electricity production needed for our computing resources --
estimated to be 75--90\% of our electricity consumption --
particularly our use of super-computing facilities. Since large-scale
simulations will continue to be an important part of astrophysics also
in future decades, we need to identify effective measures to reduce
the associated emissions. Note that we did not assess the emissions
associated with the manufacture of cluster hardware, only their use.

As is evident from the comparison of the computing-related emissions
from the MPIA versus Australian astronomers, the source of electricity
generation has the greatest impact on the computing-related carbon
footprint.  Thus, it is imperative that super-computing facilities be
run with renewable energy, and that the electricity required for
cooling is minimised. The sources of national/regional energy
production are decided at a political level, but the astronomical
community, and indeed individual citizens, can collectively {campaign}
for this change. As a mid-term option, super-computing facilities may
be moved to locations where renewables are available and less
electrical energy is needed for cooling, for example, to Iceland,
which has an average of 0.028\,kg\COT eq/kWh
emission\cite{electricitymap} for produced electricity in August
2020. Additionally, potential idle times, and hence the required
amount of hardware, could be reduced by switching to more cloud
computing, because there, capacity utilization is generally higher
than for local computers\cite{Radu_2017}. As a community, we should
guarantee an efficient use of super-computing resources. This applies
both to code efficiency, as well as regarding the computing
architecture that we build up or
rent\cite{2020arXiv200205651H,computing2020}. All these options will
require changes at the institutional and astronomy-community level.

\subsection*{Heating and local energy production}

Finally, we briefly touch on the MPIA's buildings. The use of oil for
heating at 446\,\tceq\ is the third largest contributor to the MPIA's
GHG emissions.  Oil has been used since the institute's buildings were
inaugurated in 1976, due to their distance from the city's district
heating and gas network.  For the future, the only viable and
sustainable option for heating the institute is to use ground heat, in
combination with an electrically-operated heat pump. This type of
heating system is already employed at the House of Astronomy, the
astronomy education and outreach center built on the MPIA campus in
2011. Not only can this heating system save 50\% of energy compared to
oil/gas-based systems, but also it can be run carbon-neutral on
renewable electricity. Installation of such a heating system can, in
principle, be implemented at the institute level, as can improvements
to building insulation, which reduce the heating needs. These changes
have been proposed for the MPIA and are currently under review.

MPIA's electricity is consumed both on campus for a mix of computing
(including cooling), workshops, cleanrooms, and general office
consumption, and to a large part in external high-performance
computing centers as described above.  While the associated carbon
emissions will decrease along with Germany's decreasing use of coal
and gas for electricity generation, this process will take a long
time. We note that the MPIA's utility contracts have a carbon
intensity about half that of the German average, and in principle, for
a relatively small extra cost, these contracts could be changed to
provide 100\% renewable electricity. However, many such contracts
would not actually lead to more renewable energy being produced, but
instead, only formally redistribute renewable electricity volumes or
emission certificates between contracts. Thus, in reality other
measures would have a greater de facto impact. We proposed a
photovoltaic installation on MPIA's roof, also currently under review,
which would initially produce $\sim$10\% of MPIA's on-campus
electricity consumption at zero additional cost.

\section*{Conclusion}

We have assessed and summarised the MPIA's research-related emissions
for the year 2018, finding that the average MPIA astronomer produced
at least 18.1 \tceq\ of research-related GHG emissions in that year,
{a sobering 3 times} the emissions needed for Germany to meet its 2030
goals set in accordance with the Paris agreement. We identified the
areas in which we produced the most GHG emissions and urge other
institutes to conduct their own assessment. Each institute will face a
unique set of challenges, depending on its location, funding
structure, and fields of research. These challenges can only be
addressed once quantified\cite{canada2019}. However, many of the
challenges will overlap, as is apparent from our comparison of the
MPIA's emissions to those of the Australian astronomical community.

We identified a high carbon cost associated with astronomy-wide
issues, but also a few that were institute specific. Overall,
work-related travel dominates our carbon footprint and must be
addressed as a community. If we continue to travel by air as we do
now, we will not meet the required global reduction in
\COT\ emissions. The astronomical community should adopt some of the
recommendations of Klöwer et al.\cite{kloewer2020}, and go beyond them
in some respects. The second dominant contribution is the electricity
generation for computations on clusters. Changes in the production of
electricity are required to address this in the long term, but we can
start to partially address this at the institute level with on-site,
renewable means of energy production. For example, we have proposed
the installation of solar panels on the flat and vacant MPIA roof
space.  The third highest contribution, which was institute-specific,
was the high carbon footprint associated with heating. We have
recommended that the heating system be changed to a ground heating
system in the future.

We require both a local and community-wide approach to reduce the GHG
emissions associated with astronomy research. For this, we need the
lead of both our professional organisations (e.g.\ International
Astronomical Union, European and American Astronomical Societies) and
funding agencies, as well as the development and leading by example of
larger institutes or communities. The political landscape is unlikely
to adapt rapidly enough to the evolving climate crisis
situation. Instead, {we as astronomers need to `own' our emissions and
  adapt the culture and technology we use to conduct our research.} In
doing so, we can set an example for others to follow.


\begin{thebibliography}{10}
\urlstyle{rm}
\expandafter\ifx\csname url\endcsname\relax
  \def\url#1{\texttt{#1}}\fi
\expandafter\ifx\csname urlprefix\endcsname\relax\def\urlprefix{URL }\fi
\expandafter\ifx\csname doiprefix\endcsname\relax\def\doiprefix{DOI: }\fi
\providecommand{\bibinfo}[2]{#2}
\providecommand{\eprint}[2][]{\url{#2}}

\bibitem{WHO_2003}
\bibinfo{author}{McMichael, A.} \emph{et~al.}
\newblock \emph{\bibinfo{title}{Climate change and human health: Risks and
  Responses}} (\bibinfo{publisher}{{World Health Organisation}},
  \bibinfo{year}{2003}).

\bibitem{IPCC_policymakers}
\bibinfo{author}{Edenhofer, O.} \emph{et~al.}
\newblock \bibinfo{title}{{IPCC, 2014: Summary for Policymakers.}}
  (\bibinfo{year}{2014}).

\bibitem{berkeleyearth2019}
\bibinfo{author}{Rohde, R.}
\newblock \bibinfo{title}{{Global Temperature Report for 2019}}.
\newblock
  \bibinfo{howpublished}{http://berkeleyearth.org/2019-temperatures-new/}.
\newblock \bibinfo{note}{Accessed: 2020-08-03}.

\bibitem{parisagreement}
\bibinfo{author}{{United Nations}}.
\newblock \bibinfo{title}{{The Paris Agreement}}.
\newblock
  \bibinfo{howpublished}{https://unfccc.int/process-and-meetings/the-paris-agreement/the-paris-agreement}.
\newblock \bibinfo{note}{Accessed: 2020-08-03}.

\bibitem{ausastr}
\bibinfo{author}{Stevens, A. R.~H.}, \bibinfo{author}{Bellstedt, S.},
  \bibinfo{author}{Elahi, P.~J.} \& \bibinfo{author}{Murphy, M.~T.}
\newblock \bibinfo{journal}{\bibinfo{title}{The imperative to reduce carbon
  emissions in astronomy}}.
\newblock {\emph{\JournalTitle{Nature Astronomy}}}
\textbf{\bibinfo{volume}{4}}, \bibinfo{pages}{843--851}
(\bibinfo{year}{2020}).

\bibitem{cfht2020}
\bibinfo{author}{Flagey, N.}, \bibinfo{author}{Thronas, K.},
\bibinfo{author}{Petric, A.}, \bibinfo{author}{Withington, K.} \&
\bibinfo{author}{Seidel, M.~J.}
\newblock \bibinfo{journal}{\bibinfo{title}{Measuring carbon emissions
    at the Canada-France-Hawaii Telescope}}.
\newblock {\emph{\JournalTitle{Nature Astronomy}}}
\textbf{\bibinfo{volume}{4}}, \bibinfo{pages}{816--818}
(\bibinfo{year}{2020}).

  
\bibitem{ewass2019}
\bibinfo{author}{Burtscher, L.} \emph{et~al.}
\newblock \bibinfo{journal}{\bibinfo{title}{The carbon footprint of
    large astronomy meetings}}. 
\newblock {\emph{\JournalTitle{Nature Astronomy}}}
\textbf{\bibinfo{volume}{4}}, \bibinfo{pages}{823--825}
(\bibinfo{year}{2020}).

\bibitem{marshall2009}
\bibinfo{author}{{Marshall}, P.~J.} \emph{et~al.}
\newblock \bibinfo{title}{{Low-Energy Astrophysics: Stimulating the Reduction
  of Energy Consumption in the Next Decade}}.
\newblock In \emph{\bibinfo{booktitle}{astro2010: The Astronomy and
  Astrophysics Decadal Survey}}, vol. \bibinfo{volume}{2010},
  \bibinfo{pages}{P35} (\bibinfo{year}{2009}).
\newblock \eprint{arxiv.org/abs/0903.3384}.

\bibitem{germanCO2report}
\bibinfo{author}{{German Government, Umweltbundesamt}}.
\newblock \bibinfo{journal}{\bibinfo{title}{{Submission under the United
  Nations Framework Convention on Climate Change and the Kyoto Protocol 2020,
  National Inventory Report for the German Greenhouse Gas Inventory
  1990--2018}}}.
\newblock {\emph{\JournalTitle{Climate Change}}}
  \textbf{\bibinfo{volume}{23/2020}} (\bibinfo{year}{2020}).
\newblock
  \bibinfo{note}{{https://www.umweltbundesamt.de/publikationen/submission-under-the-united-nations-framework-5}}.

\bibitem{bundesklimaschutzgesetz}
\bibinfo{author}{{German Government}}.
\newblock \bibinfo{title}{{Gesetz zur Einführung eines
  Bundes-Klimaschutzgesetzes und zur Änderung weiterer Vorschriften}}.
\newblock
  \bibinfo{howpublished}{https://www.bmu.de/gesetz/bundes-klimaschutzgesetz}
  (\bibinfo{year}{2019}).
\newblock \bibinfo{note}{Accessed: 2020-07-30}.

\bibitem{destatis1990}
\bibinfo{author}{{German Government, Statistisches Bundesamt}}.
\newblock \bibinfo{title}{{Rückgerechnete und fortgeschriebene Bevölkerung
  auf Grundlage des Zensus 2011--1991 bis 2011}}.
\newblock
  \bibinfo{howpublished}{https://www.destatis.de/DE/Themen/Gesellschaft-Umwelt/Bevoelkerung/Bevoelkerungsstand/Publikationen/Downloads-Bevoelkerungsstand/rueckgerechnete-bevoelkerung-5124105119004.pdf}.
\newblock \bibinfo{note}{Accessed: 2020-07-30}.

\bibitem{destatis}
\bibinfo{author}{{German Government, Statistisches Bundesamt}}.
\newblock \bibinfo{title}{Germany current population}.
\newblock
  \bibinfo{howpublished}{https://www.destatis.de/EN/Themes/Society-Environment/Population/Current-Population/\_node.html}.
\newblock \bibinfo{note}{Accessed: 2020-07-24}.

\bibitem{owidco2andothergreenhousegasemissions}
\bibinfo{author}{Ritchie, H.} \& \bibinfo{author}{Roser, M.}
\newblock \bibinfo{journal}{\bibinfo{title}{{$\mathrm{CO_2}$ and Greenhouse Gas
  Emissions}}}.
\newblock {\emph{\JournalTitle{Our World in Data}}}  (\bibinfo{year}{2017}).
\newblock
  \bibinfo{note}{{https://ourworldindata.org/co2-and-other-greenhouse-gas-emissions}}.

\bibitem{atmosfair}
\bibinfo{author}{{Atmosfair gGmbH}}.
\newblock \bibinfo{title}{Atmosfair emissions calculator}.
\newblock
  \bibinfo{howpublished}{https://www.atmosfair.de/en/standards/emissions\_calculation/
    emissions\_calculator}
  (\bibinfo{year}{2016}).
\newblock \bibinfo{note}{Accessed: 2020-07-24}.

\bibitem{qantas}
\bibinfo{title}{Qantas future planet}.
\newblock \bibinfo{howpublished}{https://www.qantasfutureplanet.com.au}.

\bibitem{Aus_energy_stats}
\bibinfo{author}{{Australian Government}}.
\newblock \bibinfo{title}{{Australian electricity generation, by fuel type,
  physical units}}.
\newblock \bibinfo{type}{Tech. Rep.}, \bibinfo{institution}{{Department of the
  Environment and Energy}} (\bibinfo{year}{2019}).

\bibitem{fraunhoferISE}
\bibinfo{author}{{Fraunhofer ISE}}.
\newblock \bibinfo{title}{{Net public electricity generation in Germany in
  2018}}.
\newblock
  \bibinfo{howpublished}{https://www.energy-charts.de/energy\_pie.htm?year=2018}.
\newblock \bibinfo{note}{Accessed: 2020-07-24}.

\bibitem{brkg}
\bibinfo{author}{{German Government, BMI}}.
\newblock \bibinfo{title}{{Beitrag zum Klimaschutz: Mehr Dienstreisen mit der
  Bahn}}.
\newblock
  \bibinfo{howpublished}{https://www.bmi.bund.de/SharedDocs/kurzmeldungen/DE/2020/01/brkg-bahn.html}.
\newblock \bibinfo{note}{Accessed: 2020-08-03}.

\bibitem{SFF1000}
\bibinfo{author}{{Scientists for Future}}.
\newblock \bibinfo{title}{{Voluntary commitment to refrain from short-haul
  business flights}}.
\newblock \bibinfo{howpublished}{{https://unter1000.scientists4future.org}}.
\newblock \bibinfo{note}{{Accessed: 2020-08-03}}.

\bibitem{kloewer2020}
\bibinfo{author}{Kl\"ower, M.}, \bibinfo{author}{Hopkins, D.},
  \bibinfo{author}{Allen, M.} \& \bibinfo{author}{Higham, J.}
\newblock \bibinfo{journal}{\bibinfo{title}{An analysis of ways to decarbonize
  conference travel after $\mathrm{COVID}$-19}}.
\newblock {\emph{\JournalTitle{Nature}}} \textbf{\bibinfo{volume}{583}},
  \bibinfo{pages}{356--359} (\bibinfo{year}{2020}).

\bibitem{electricitymap}
\bibinfo{howpublished}{https://www.electricitymap.org}.
\newblock \bibinfo{note}{Accessed: 2020-08-03}.

\bibitem{Radu_2017}
\bibinfo{author}{Radu, L.-D.}
\newblock \bibinfo{journal}{\bibinfo{title}{Green cloud computing: A literature
  survey}}.
\newblock {\emph{\JournalTitle{Symmetry}}} \textbf{\bibinfo{volume}{9}},
  \bibinfo{pages}{295}, \doiprefix\url{10.3390/sym9120295}
  (\bibinfo{year}{2017}).

\bibitem{2020arXiv200205651H}
\bibinfo{author}{{Henderson}, P.} \emph{et~al.}
\newblock \bibinfo{journal}{\bibinfo{title}{{Towards the Systematic Reporting
  of the Energy and Carbon Footprints of Machine Learning}}}.
\newblock {\emph{\JournalTitle{arXiv e-prints}}}
  \bibinfo{pages}{arXiv:2002.05651} (\bibinfo{year}{2020}).
\newblock \eprint{2002.05651}.

\bibitem{computing2020}
\bibinfo{author}{Portegies Zwart, S.}
\newblock \bibinfo{journal}{\bibinfo{title}{The ecological impact of high-performance computing in astrophysics}}.
\newblock {\emph{\JournalTitle{Nature Astronomy}}}
\textbf{\bibinfo{volume}{4}}, \bibinfo{pages}{819--822}
(\bibinfo{year}{2020}).

\bibitem{canada2019}
\bibinfo{author}{{Matzner}, C.} \emph{et~al.}
\newblock \bibinfo{title}{{Astronomy in a Low-Carbon Future}}.
\newblock In \emph{\bibinfo{booktitle}{Canadian Long Range Plan for Astronony
  and Astrophysics White Papers}}, vol. \bibinfo{volume}{2020},
  \bibinfo{pages}{22}, \doiprefix\url{10.5281/zenodo.3758549}
  (\bibinfo{year}{2019}).
\newblock \eprint{1910.01272}.


\end{thebibliography}


\section*{Additional information}
The authors declare no competing interests.

\end{document}